\documentclass[aps,pra,preprint]{revtex4-1}
\usepackage{hyperref}
\usepackage{amsmath}
\usepackage{amssymb}
\usepackage[pdftex]{graphicx}
\usepackage{gentium}
\usepackage{microtype}
\usepackage{verbatim}

\makeatletter
\DeclareRobustCommand{\element}[1]{\@element#1\@nil}
\def\@element#1#2\@nil{%
  #1%
  \if\relax#2\relax\else\MakeLowercase{#2}\fi}
\makeatother

\newcommand{\mrm}{\mathrm}

\newcommand{\id}{\mathbb{I}}
\newcommand{\ket}[1]{{\left| {#1} \right\rangle}}

\newcommand{\braket}[2]{{\left\langle {#1}|{#2} \right\rangle}}
\newcommand{\avg}[1]{\left\langle {#1} \right\rangle }

\begin{document}
\title{The collisional frequency shift of a trapped-ion optical clock}
\author{Amar C. Vutha}
\email{vutha@physics.utoronto.ca}
\affiliation{Department of Physics, University of Toronto, 60 St.\ George Street, Toronto, Canada M5S 1A7}
\author{Tom Kirchner}
\affiliation{Department of Physics and Astronomy, York University, 4700 Keele Street, Toronto, Canada M3J 1P3}
\author{Pierre Dub\'{e}}
\affiliation{National Research Council Canada, Ottawa, Canada K1A 0R6}

\begin{abstract}
Collisions with background gas can perturb the transition frequency of trapped ions in an optical atomic clock. We develop a non-perturbative framework based on a quantum channel description of the scattering process, and use it to derive a master equation which leads to a simple analytic expression for the collisional frequency shift. As a demonstration of our method, we calculate the frequency shift of the Sr$^+$ optical atomic clock transition due to elastic collisions with helium.
\end{abstract}
\maketitle

\section{Introduction}
Optical clocks using trapped ions have greatly improved in precision and accuracy over the last decade \cite{Chou2010, King2012, Madej2012, Margolis2014, Barwood2014, Dube2014, Ludlow2015, Huntemann2016}. In optical clocks that use trapped atomic ions, a potentially important contribution to the systematic error budget is the collisional frequency shift (CFS), which arises due to collisions of the clock ion with residual gas atoms and molecules in the ion trap vacuum chamber. Collisional frequency shifts have previously been estimated for cold neutral atom clocks in \cite{Gibble2013}, under the assumption that large-angle collisions transfer enough momentum to remove the atoms from their trap and can therefore be neglected -- however, this assumption is not valid for ion traps, where the trap depths can be much larger than in neutral atom traps. Mitroy \emph{et al.} \cite{Mitroy2008} calculated some dispersion coefficients for Sr$^+$ ions colliding with buffer gas atoms, but the implicit assumption of long-range interactions is not justified as low-partial-wave collisions are sensitive to the full range of the inter-atomic potential. For ion clocks, worst-case estimates were provided in \cite{Rosenband2008}, under the assumption that each collision between the clock ion and a background gas molecule imprints the largest possible phase shift between the ground and excited states of the ion. For the Al$^{+}$ ion, this worst-case scenario gives an estimated fractional frequency shift uncertainty of $0.5\times 10^{-18}$~\cite{Rosenband2008}, and for the Sr$^{+}$ ion $2\times 10^{-18}$~\cite{Dube2013}. 

With rapid improvements in the control of the systematic shifts towards the $10^{-18}$ level, better knowledge of the CFS becomes essential for obtaining ultimate clock performance. For the particular case of the Sr$^{+}$ ion, it is anticipated that progress in the evaluation of the blackbody radiation field environment~\cite{Dolezal2015}, and the use of specially designed ion traps for low systematic shifts~\cite{NisbetJones2016}, will bring the total uncertainty down to the level of the collisional shift in the near future~\cite{Dube2013, Dube2014}. 

In this paper, we develop a systematic procedure to evaluate the CFS. Our method provides a generally applicable alternative to perturbative analyses of the scattering process, which can be unreliable when collisions impart large phase shifts to the clock ions (as is often the case for short-range collisions). In Section \ref{sec:formalism}, we obtain a master equation to describe collisional shifts, which we develop into a simple expression for the CFS of a two-level clock atom in Section \ref{sec:two_level_formalism}. Following this, in Section \ref{sec:numerical}, we evaluate the scattering phase shifts using \emph{ab initio} potential curves for the SrHe$^+$ molecule, and use these phase shifts to estimate the CFS of a Sr$^+$ clock due to collisions with helium gas. 

\section{General formalism}\label{sec:formalism}
Consider a clock ion with $n$ energy levels, labeled $\ket{\alpha}, (\alpha = 0,1,2 \ldots n)$, which undergoes elastic collisions with background gas atoms. We assume that the background gas atoms have only a single energy level that is accessible at the energies involved in the collision, which is a good model for background gas atoms such as helium. We obtain the scattered wavefunction in the usual way, by calculating the eigenfunctions of the Schr\"odinger equation. The combined wavefunction representing the relative motional state of the clock ion and gas atom, and the vector of internal states of the ion, is denoted as $\Psi$. The Schr\"odinger equation is written as 
\begin{equation}
\left(-\frac{\hbar^2}{2\mu} \nabla^2 \id + \mathbb{V}(r) \right) \Psi = \frac{\hbar^2 k^2}{2\mu} \Psi,
\end{equation}
where $\mu$ is the reduced mass of the clock ion and background gas atom, $r$ denotes their separation relative to their center of mass, $v = \hbar k/\mu$ is the relative velocity of the colliding partners, and the elements of the diagonal matrix $\mathbb{V}(r)$ are the potential energy curves $V_\alpha(r)$ for the ion-atom interaction that connect adiabatically and asymptotically to the ion's internal eigenstates $\ket{\alpha}$, i.e., $[\mathbb{V}]_{\alpha \beta} = V_\alpha(r) \, \delta_{\alpha \beta}$. We denote the incoming internal state of the ion as $\psi_\mrm{in}$. As $r \to \infty$, the eigenfunction $\Psi$ can be written as 
\begin{equation}
\Psi(\vec{r}) = \mathcal{A} \, e^{i \vec{k}\cdot\vec{r}} \, \psi_\mrm{in}  + \Psi_s(r,\theta),
\end{equation}
where the scattered wavefunction $\Psi_s$ is written using the standard partial-wave expansion as
\begin{equation}\label{eq:scattered_wavefunction}
\Psi_s(r,\theta) = \mathcal{A} \left[\sum_{\ell=0}^\infty \frac{\sqrt{4\pi(2\ell + 1)}}{k} \frac{e^{i kr}}{r} Y_{\ell 0}(\theta) \, \mathbb{T}_\ell \right] \psi_\mrm{in}.
\end{equation}
Here $\mathcal{A}$ is an arbitrary normalization factor, and $\mathbb{T}_\ell$ is the $n$-dimensional $T$-matrix whose entries are $[\mathbb{T}_\ell]_{\alpha \beta} = \sin \phi_{\ell,\alpha} \, e^{i \phi_{\ell,\alpha}} \, \delta_{\alpha \beta}$. The quantity $\phi_{\ell,\alpha}$ is the scattering phase shift for the $\ell$-th partial wave on the potential energy curve $V_\alpha(r)$. In writing the $T$-matrix as a diagonal matrix for the internal states of the ion, we have made the assumption that the adiabatic potential energy curves of the ion-atom system do not interact at any value of $r$, and therefore that the scattering process does not cause transitions between the internal states of the ion.
%
%

We will construct a master equation to model the dynamics of the reduced density matrix of the ion, over time intervals $\delta t \gg \tau_s$, where $\tau_s \sim \mu r_\alpha/\hbar k$ is the duration of a scattering event ($r_\alpha$ being the range of the potential $V_\alpha(r)$). We begin by associating an $n$-dimensional Kraus operator $M_\ell$ with each partial wave $\ell$, which is obtained by projecting the scattered wavefunction (the joint internal-motional wavefunction) onto an outgoing spherical wave basis state (for the motional degree of freedom). For cylindrically symmetric problems, the functions $g_\ell(r,\theta) = \frac{e^{ikr}}{r} Y_{\ell 0}(\theta)$ provide an orthonormal basis set for the outgoing spherical waves with regard to integration on the surface of a sphere. The Kraus operator $M_\ell$ is defined such that
\begin{equation}
M_\ell \psi_\mrm{in} \equiv \int r^2 d\Omega \, g_\ell^*(r,\theta) \, \Psi_s(r,\theta).
\end{equation}
Using Equation \eqref{eq:scattered_wavefunction}, $M_\ell$ can be related to the $T$-matrix as
\begin{equation}
M_\ell = \mathcal{A} \frac{\sqrt{4\pi(2\ell + 1)}}{k} \mathbb{T}_\ell.
\end{equation}
We choose the normalization factor so that $|\mathcal{A}|^2$ equals the number of background gas atoms, per unit area, that are encountered by the clock ion in a duration $\delta t$. Hence $\mathcal{A} = \sqrt{n v \delta t}$, where $n$ is the number density of the background gas.

With this choice of $\mathcal{A}$, the matrix elements of the Kraus operators are $[M_\ell]_{\alpha \beta} = \sqrt{\gamma_{\ell,\alpha} \delta t} \, e^{i \phi_{\ell,\alpha}} \, \delta_{\alpha \beta}$. The partial wave scattering rates are $\gamma_{\ell,\alpha} = n v \sigma_{\ell,\alpha}$, where $\sigma_{\ell,\alpha} = \frac{4 \pi}{k^2} (2\ell+1) \sin^2\phi_{\ell,\alpha}$ is the $\ell$-th partial wave cross-section on the potential curve $V_\alpha(r)$.

Using these Kraus operators, the time evolution of the reduced density matrix of the ion over a duration $\delta t$ can be written as a quantum channel that has the operator sum expansion
\begin{equation}\label{eq:operator_sum}
\rho(\delta t) = M_\emptyset \rho(0) M_\emptyset^\dagger + \sum_{\ell=0}^\infty M_\ell \rho(0) M_\ell^\dagger.
\end{equation}
The operator $M_\emptyset$ is chosen so that the completeness relation $M_\emptyset^\dagger M_\emptyset + \sum_{\ell=0}^\infty M_\ell^\dagger M_\ell = \id$ is satisfied, which ensures that the transformation $\rho(0) \to \rho(\delta t)$ is trace-preserving. The matrix elements of $M_\emptyset$ are
\begin{equation}
[M_\emptyset]_{\alpha \beta} = \sqrt{1-p_\alpha} \, \delta_{\alpha \beta},
\end{equation}
where $p_\alpha = \sum_{\ell=0}^\infty \gamma_{\ell,\alpha} \delta t = n \sigma_\alpha v \delta t$ is the scattering probability, and $\sigma_\alpha = \sum_\ell \sigma_{\ell,\alpha}$ is the total scattering cross-section, for the potential curve $V_\alpha(r)$. The operator $M_\emptyset$ can be sensibly defined when the time step is sufficiently small, $\delta t \ll 1/n \sigma_\alpha v$. For small values of $\delta t$, $M_\emptyset \approx \id -\frac{1}{2} \sum_\ell M_\ell^\dagger M_\ell + o(\delta t)$. 

Therefore, for small time intervals $\delta t$ (but large enough that $\delta t \gg \tau_s$ remains valid), the transformation $\rho(0) \to \rho(\delta t)$ in \eqref{eq:operator_sum} can be turned into a first-order differential equation in Lindblad form, under the assumption that the bath of background gas atoms has the Markov property. The Markovian assumption is appropriate here, since the trajectories of the background gas atoms are randomized after collisions with, e.g., the walls of the vacuum chamber around the clock ion. Hence it is safe to assume that consecutive collisions between the clock ion and background gas atoms are completely uncorrelated. The Lindblad equation for the density matrix of the internal states of the ion is therefore
\begin{equation}
\frac{d \rho}{d t} = -i [H,\rho] + \sum_\ell L_\ell \rho L_\ell^\dagger - \frac{1}{2}\sum_\ell \left\{ L_\ell^\dagger L_\ell, \rho \right\}.
\end{equation}
We have included a term $-i[H,\rho]$ for the unitary evolution of the clock ion, where $H$ represents the interaction Hamiltonian of the ion with laser fields and trapping fields. The Lindblad operators are $L_\ell = M_\ell/\sqrt{\delta t}$, and their matrix elements are
\begin{equation}
[L_\ell]_{\alpha \beta} = \sqrt{\gamma_{\ell,\alpha}} \, e^{i \phi_{\ell,\alpha}} \, \delta_{\alpha \beta}.
\end{equation}

\section{Two-level clock ion}\label{sec:two_level_formalism}
We now specialize to the case when the clock ion has only two energy levels $\ket{g}, \ket{e}$, so that the index $\alpha$ runs over just $g, e$. The Lindblad operators are 
\begin{equation}
L_\ell = \begin{pmatrix} \sqrt{\gamma_{\ell,g}}\, e^{i \phi_{\ell,g}} & 0 \\ 0 & \sqrt{\gamma_{\ell,e}}\, e^{i \phi_{\ell,e}} \end{pmatrix}.
\end{equation}
The Hamiltonian of the clock ion (under the rotating wave approximation) is
\begin{equation}
H = \frac{1}{2}\begin{pmatrix}
\Delta & \Omega(t) \\ \Omega(t) & -\Delta
\end{pmatrix},
\end{equation}
where $\Omega(t)$ is the Rabi frequency for the laser-atom interaction, and $\Delta = \omega_L - \omega_0$ is the detuning of the probe laser frequency, $\omega_L$, from the atomic resonance, $\omega_0$. Using this Hamiltonian, and the Lindblad operators $L_\ell$, the master equation for the density matrix is 
\begin{equation}\label{eq:master}
\begin{split}
\frac{d\rho}{dt} & = -i \begin{pmatrix}
\frac{\Omega}{2}(\rho_{eg}-\rho_{ge}) & \frac{\Omega}{2}(\rho_{ee}-\rho_{gg}) + \Delta \,  \rho_{ge} \\
-\frac{\Omega}{2}(\rho_{ee}-\rho_{gg}) - \Delta \, \rho_{eg} & -\frac{\Omega}{2}(\rho_{eg}-\rho_{ge})
\end{pmatrix} + \sum_\ell \begin{pmatrix}
0 & \kappa_\ell \, \rho_{ge} \\ \kappa_\ell^\dagger \, \rho_{eg} & 0 \end{pmatrix}.
\end{split}
\end{equation}
The coefficients $\kappa_\ell$ appearing in \eqref{eq:master} are \begin{equation}
\begin{split}
\kappa_\ell & = \sqrt{\gamma_{\ell,g} \gamma_{\ell,e}} \, e^{i \left(\phi_{\ell,g} - \phi_{\ell,e}\right) } - \bar{\gamma}_\ell  \\
& = \left[ \sqrt{\gamma_{\ell,g} \gamma_{\ell,e}} \, \cos \left(\phi_{\ell,g} - \phi_{\ell,e}\right) - \bar{\gamma}_\ell \right] + i  \sqrt{\gamma_{\ell,g} \gamma_{\ell,e}}\, \sin \left(\phi_{\ell,g} - \phi_{\ell,e}\right),  
\end{split}
\end{equation}
with $\bar{\gamma}_\ell = \frac{\gamma_{\ell,g} + \gamma_{\ell,e}}{2}$ being the average collision rate for the two clock states in the $\ell$-wave scattering channel. Note that $\mathfrak{Re} (\kappa_\ell) \leq 0$ always, regardless of the values of the scattering phase shifts and collision rates. Therefore collisions always result in the decay of the off-diagonal coherences $\rho_{ge}, \rho_{eg}$, with a damping coefficient $\Gamma_c = -\sum_\ell \mathfrak{Re} (\kappa_\ell)$, which shows up in experiments as a decrease of the contrast of Rabi oscillations or Ramsey fringes. 

Frequency shifts are related to the imaginary parts of $\kappa_\ell$, which appear \emph{on equal footing with the detuning} $\Delta = \omega_L - \omega_0$ in \eqref{eq:master}. Therefore, regardless of the details of the experimental sequence, or the values of parameters such as $\Omega(t)$ that determine the lineshape, the observed resonance frequency of the two-level clock ion should be corrected by an amount
\begin{equation}
\begin{split} \label{eq:frequency_shifts}
\delta \omega & = -\sum_\ell \mathfrak{Im}(\kappa_\ell) = -\sum_\ell \sqrt{\gamma_{\ell,g} \gamma_{\ell,e}} \, \sin \left(\phi_{\ell,g} - \phi_{\ell,e}\right).
\end{split}
\end{equation}
We hereafter refer to $\delta \omega$ as the CFS.
%
%
%
%
%
%
Inserting the explicit expressions for the partial wave scattering rates and using the standard form of the scattering amplitudes (e.g., for the ground state)
\begin{equation}
f_g(\theta) = \sum_{\ell}   \frac{\sqrt{4\pi(2\ell + 1)}}{k}  \sin\phi_{\ell,g} \, e^{i\phi_{\ell,g}} Y_{\ell 0}(\theta),
\end{equation}
it is easy to see that the CFS can be cast into the compact form
\begin{equation}\label{eq:CFS_compact}
\delta \omega = - n v \,\mathfrak{Im} \int d\Omega \, f_e^*(\theta) f_g (\theta).
\end{equation}
The CFS is determined by the overlap of the scattering amplitudes for the two clock states. This result can also be inferred from the formalism of Ref.~\cite{Hornberger2007}. Equation \eqref{eq:CFS_compact} implies that $\delta \omega = 0$ in the first Born approximation, since first Born scattering amplitudes are real for radially symmetric real potential functions such as $V_{g} (r)$ and $V_{e}(r)$. 

Furthermore, using the Cauchy-Schwarz inequality with either \eqref{eq:frequency_shifts} or \eqref{eq:CFS_compact}, we obtain the bound 
\begin{equation}\label{eq:cfs_bound}
|\delta \omega| \leq \sqrt{\gamma_{\mrm{tot},g} \, \gamma_{\mrm{tot},e}},
\end{equation}
where the total scattering rate is $\gamma_{\mrm{tot},g} = \sum_\ell \gamma_{\ell,g} = n v \int d\Omega |f_{g}(\theta)|^2$ for the ground state (with a similar expression for the excited state). For clock atoms and ions where accurate estimates for the CFS may not be available, \eqref{eq:cfs_bound} can be used to obtain conservative upper bounds for the CFS based on measured or estimated collision rates. We emphasize that all of our results are non-perturbative, and can be applied even in situations where the ions and scatterers interact strongly. 

In passing, we note that replacing both $\gamma_{\mrm{tot},g}$ and $\gamma_{\mrm{tot},e}$ in \eqref{eq:cfs_bound} by a total scattering rate $k$ leads to the bound $\frac{|\delta \omega|}{2\pi} \leq 0.16 k$, which agrees closely with the estimate provided for the CFS in Ref.~\cite{Rosenband2008}. 

Returning to Equation \eqref{eq:frequency_shifts} we note that, in principle, the summation over partial waves should only be carried out over the values of $\ell$ that lead to recoils small enough that the ion remains detectable by the state-readout laser in the atomic clock. The recoil imparted by an $\ell$-wave collision can be calculated using the partial-wave momentum transfer cross-sections (e.g., for the ground state)
\begin{equation}
\sigma_{\mrm{mt, \ell},g} = \frac{4\pi}{k^2} (\ell+1) \sin^2\left(\phi_{\ell+1,g} - \phi_{\ell,g} \right),
\end{equation}
along with the fact that the average velocity kick delivered to the ion in an $\ell$-wave collision is
\begin{equation}
\avg{\Delta v}_\ell = \frac{\hbar k}{m_\mrm{ion}} \frac{\sigma_{\mrm{mt,\ell},g}}{\sigma_{\mrm{tot},g}},
\end{equation}
where $\sigma_{\mrm{tot},g} = \frac{4 \pi}{k^2} \sum_{\ell=0}^\infty (2\ell+1) \, \sin^2\phi_{\ell,g}$ and $m_\mrm{ion}$ is the mass of the trapped ion.

For the particular problem of Sr$^+$-He collisions at room temperature that we consider in the following section, the typical recoil velocities of the Sr$^+$ ion are $<$ 10 m/s, corresponding to maximum Doppler shifts of 24 MHz that are not much larger than the $\Gamma_{sp} = 2 \pi \times$ 22 MHz natural linewidth of the $5p \, ^2P_{1/2} \to 5s \, ^2S_{1/2}$ transition ($\lambda_{sp} \approx $ 422 nm) used for detecting the state of the ion. Hence the recoils are not practically resolvable in this case, and so we conservatively include all partial waves (up to $\ell_\mrm{max}$ = 90) in the evaluation of the collisional frequency shift. In experiments where small recoils of the clock ion can be detected, it may be possible to obtain more information about the range of contributing partial waves, and thereby reduce the systematic uncertainty associated with collisions.

\section{Numerical results for \element{Sr}$^+$-\element{He} collisions} \label{sec:numerical}
As an illustration of the general method for calculating the CFS, we describe our numerical calculations for the case of Sr$^+$-He collisions in this Section. The potential energy curves for the SrHe$^+$ diatomic molecule were calculated using PSI4 \cite{Parrish2017}. For the results reported in this section, we used the \texttt{DZVP} basis set for Sr \cite{Godbout1992}, and the \texttt{cc-pVTZ} basis set for He \cite{Woon1994}. 

\begin{figure}
\centering
\includegraphics[width=\columnwidth]{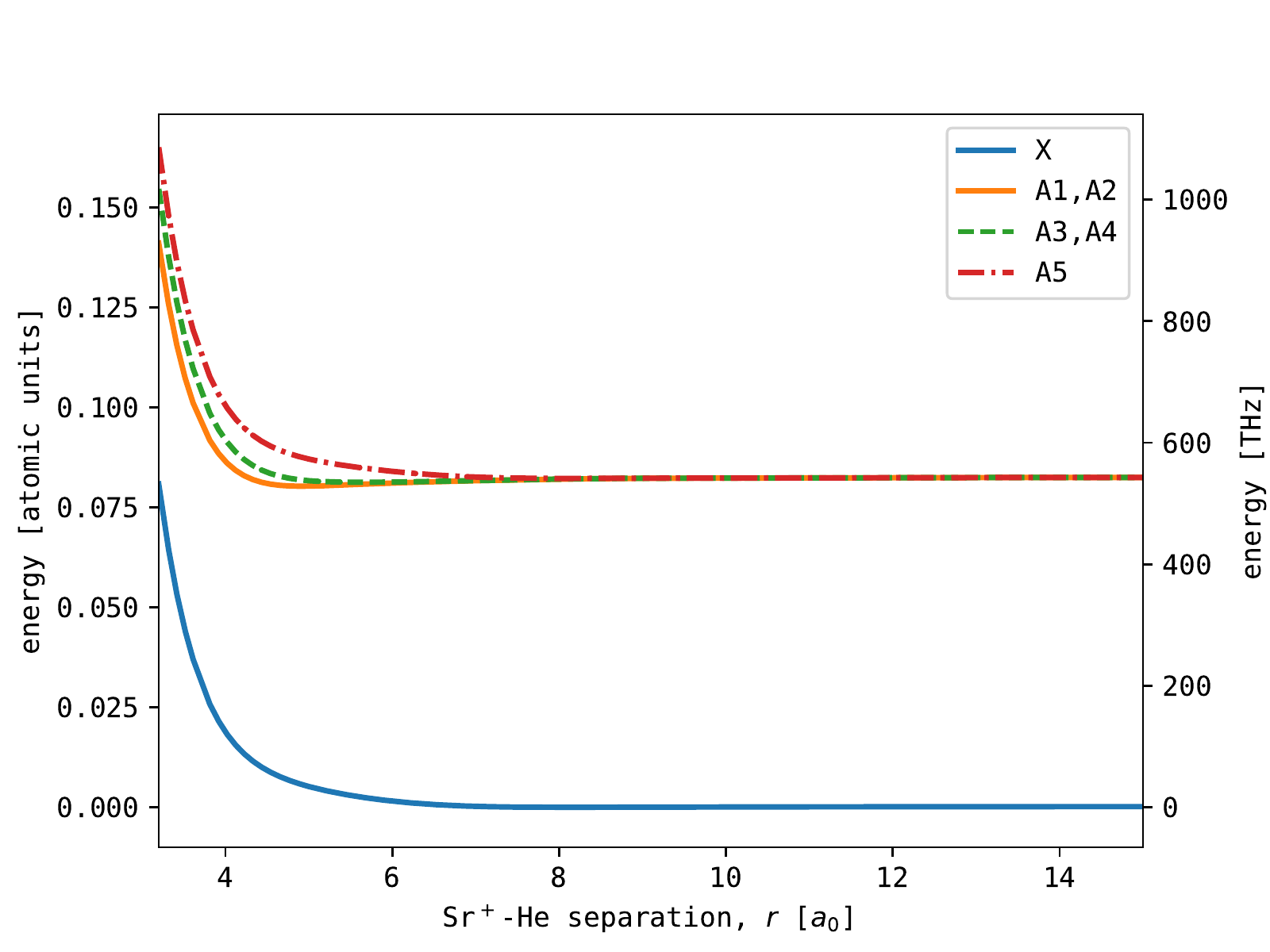}
\caption{Calculated potential energy curves for the SrHe$^+$ molecule. The curve labeled $X$ adiabatically connects to the Sr$^+$ ${}^2S_{1/2}$ state, while the five curves labeled $A1-A5$ connect to the Sr$^+$ ${}^2D_J$ states. Our calculations neglect the fine structure splitting of the ${}^2D_J$ states.}
\label{fig:potential_curves}
\end{figure}

The potential energy curves (PECs), for the ground state and a number of excited states, were calculated using the equation of motion coupled cluster (EOM-CCSD) method \cite{Wang2014} as implemented in the PSI4 package. The EOM-CCSD eigen-energies were calculated for Sr$^+$-He separations ranging from 3-30 $a_0$ to obtain the adiabatic PECs. The low-lying PECs that are relevant to the Sr$^+$-He collisional frequency shift calculation are shown in Figure \ref{fig:potential_curves}. 

The calculated PECs that asymptotically connect to the Sr$^+$ $5s \, {}^2S_{1/2}$ and $4d \, {}^2D_{J}$ states (hereafter referred to as $S$ and $D$ states, respectively) were then used as inputs for a calculation of the partial-wave scattering phase shifts. Accurately accounting for spin-orbit interactions in excited-state potential energy curves is a difficult problem in general (cf. \cite{Mitroy2008}), and our PEC calculations do not include spin-orbit coupling in the $D$ states of the Sr$^+$ ion. We find that the \emph{ab initio} PECs yield a $D$ state energy for the Sr$^+$ ion that differs from the correct value (445 THz) by approximately 100 THz. We attribute this discrepancy to the neglect of the spin-orbit interaction in the $D$ states --  however, we note that the scattering phase shift along a PEC is not affected by the addition of an overall constant asymptotic energy, but instead depends on the detailed shape of the PEC. 

\begin{figure}
\centering
\includegraphics[width=\columnwidth]{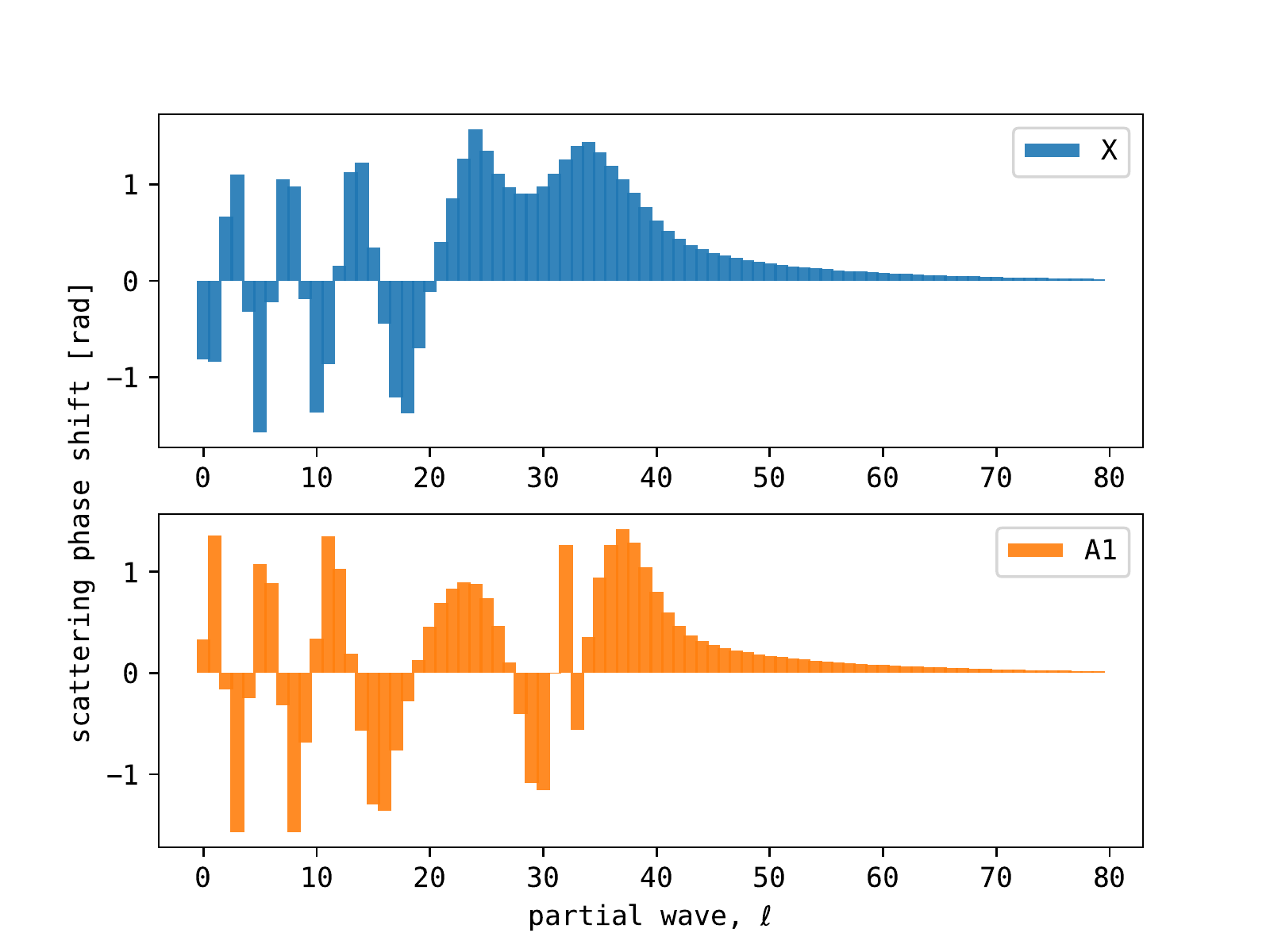}
\caption{An example of calculated scattering phase shifts at a collision energy of 290 K. The phase shifts are shown as a function of the partial wave angular momentum, $\ell$, for collisions of He with Sr$^+$ in the ground $S$ (excited $D$) state, corresponding to the $X$ ($A1$) molecular potential energy curve of SrHe$^+$.}
\label{fig:phaseshifts}
\end{figure}

\begin{figure}
\centering
\includegraphics[width=\columnwidth]{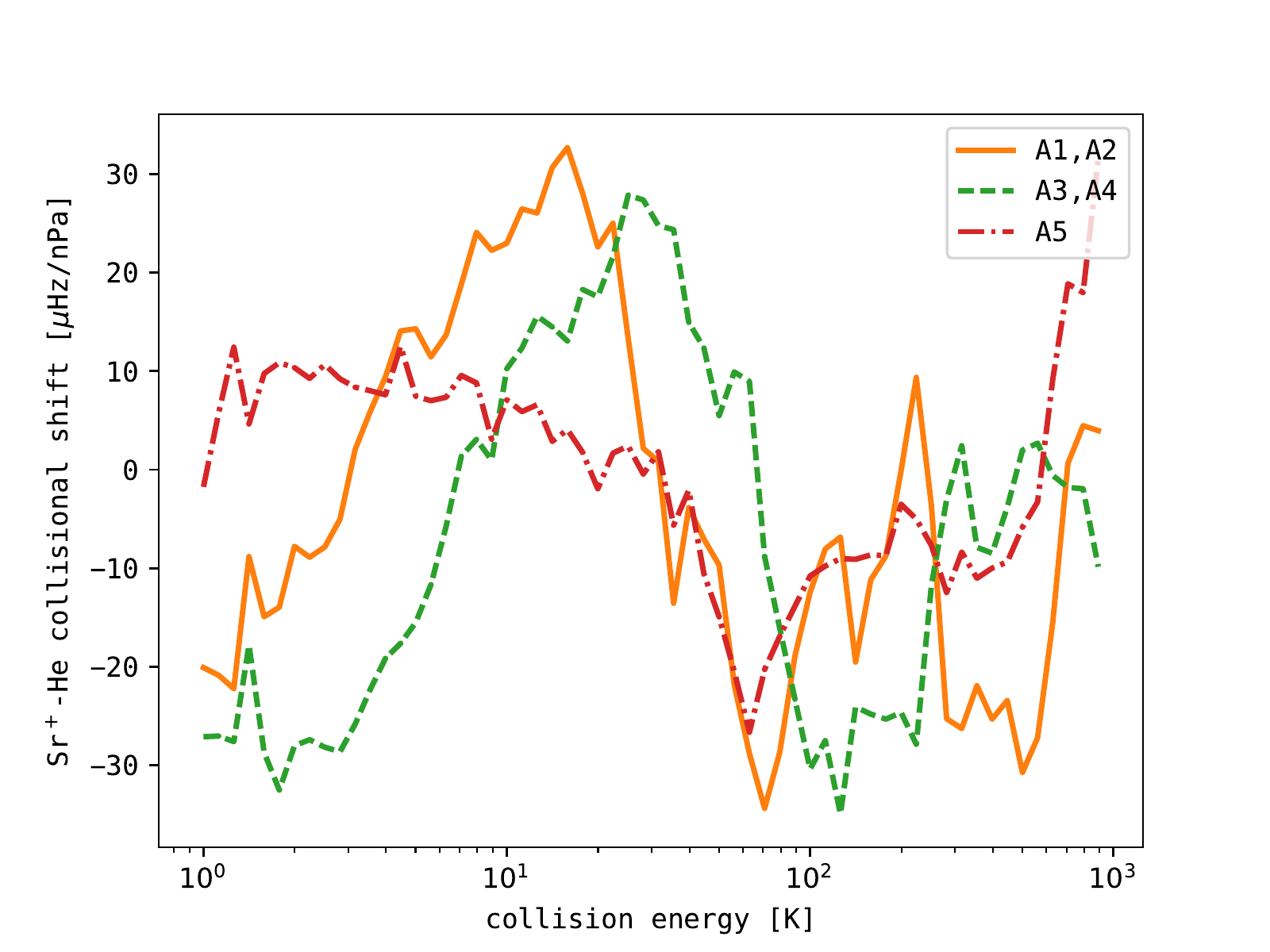}
\caption{Collisional frequency shifts of the clock transition as a function of the kinetic energy (in units of kelvin) of the colliding partners. The frequency shift was calculated using partial-wave phase shifts, such as those shown in Figure \ref{fig:phaseshifts}, along with Equation \eqref{eq:frequency_shifts}. The contributions of each of the PECs that asymptote to the $D$ state are shown separately.}
\label{fig:cfs_versus_energy}
\end{figure}

For the scattering calculations, we used the Numerov method to integrate the radial Schr\"odinger equation, and obtained the scattering phase shifts for partial waves from $\ell=0$ to 90. Examples of the calculated phase shifts are shown in Figure \ref{fig:phaseshifts}. The scattering phase shifts were then used to obtain the CFS, using \eqref{eq:frequency_shifts}. We verified that the inequality in \eqref{eq:cfs_bound} was satisfied by all of our numerical results. 

To check our results further, we compared our numerical calculations against the asymptotic formula for ion-atom collision cross sections
\begin{equation}\label{eq:asymptotic_cross_section}
\sigma_\mrm{tot,asymp} = \pi \left( 1 + \frac{\pi^2}{16}\right) \left(\frac{\mu C_4^2}{\hbar^2 E} \right)^{1/3},
\end{equation}
where $C_4$ is the dipole polarizability of helium and $E$ is the collision energy. This equation was derived in Ref.~\cite{Cote2000} under the assumption that the collision is dominated by the $\frac{1}{r^4}$ long-range interaction between the ion and the polarizable atom (and is independent of the internal state of the ion). Our numerically calculated cross sections agreed well with this asymptotic formula at high temperatures ($\gtrsim$600~K), but larger discrepancies were observed at lower temperatures, as expected. For example, at 290~K, we calculated the total scattering cross section to be $\sigma_{\mrm{tot},S} = 2.4 \times 10^{-14}$ cm$^2$ for the $S$ state, and $\avg{\sigma_{\mrm{tot},D}} = 2.9 \times 10^{-14}$ cm$^2$ averaged over the $D$ state PECs. In comparison, \eqref{eq:asymptotic_cross_section} leads to the value $\sigma_\mrm{tot,asymp} = 3.5 \times 10^{-14}$ cm$^2$.

The calculation of the CFS was repeated for collision energies ranging from 1-1000~K, with the results shown in Figure \ref{fig:cfs_versus_energy}. These CFS values were averaged with weights drawn from a Boltzmann distribution corresponding to the temperature $T$ = 290~K, in order to estimate the CFS, $\delta \omega_n$, for each of the five $D$ state potential energy curves $An$ ($n=1$ to $5$). These values of $\delta \omega_n$ were averaged together to obtain the final estimate $\avg{\delta \omega}$. We conservatively assign twice the maximum difference between $\avg{\delta \omega}$ and the individual values $\delta \omega_n$ as the uncertainty in $\avg{\delta \omega}$. The differences between $\delta \omega_n$ also provide a conservative indication of the uncertainty due to errors in the \textit{ab initio} PEC calculations. 

The estimated CFS at 290 K is $\avg{\delta \omega} = 2\pi \times (-2.6 \pm 8.5)$~$\mu$Hz/nPa. In order to test the sensitivity of this result to the choice of basis set used for the \emph{ab initio} PECs, we repeated the entire set of calculations using the \texttt{TZP} basis set for Sr \cite{Campos2013}. While this basis set is nominally of a higher quality than \texttt{DZVP}, it yielded an asymptotic $D$ state energy that was $\sim$200 THz higher than the correct value. The collisional shift obtained with this basis set was $\avg{\delta \omega}_\mrm{TZP} = 2\pi \times (-0.1 \pm 6.9)$~$\mu$Hz/nPa, which agrees with the result from the \texttt{DZVP} basis set to within its uncertainty. Since a more accurate $D$ state energy was obtained with \texttt{DZVP} basis set however, Figures \ref{fig:potential_curves}--\ref{fig:cfs_versus_energy} and the final results reported here are based on the \texttt{DZVP} potential energy curves.

The value $\avg{\delta \omega} = 2\pi \times (-2.6 \pm 8.5)$~$\mu$Hz/nPa corresponds to a fractional frequency shift $\frac{\avg{\delta \omega}}{\omega_0} = (-6 \pm 19) \times 10^{-21}$/nPa for the $\omega_0 = 2 \pi \times$ 444.8 THz clock transition in Sr$^+$ due to collisions with helium. Despite the uncertainty, this is a significantly improved (and smaller) estimate of the collisional shift systematic error compared to the worst-case estimate in Ref.\ \cite{Dube2013}. 

Finally, we also studied the variation in $\avg{\delta \omega}$ when the temperature $T$ was changed from 200~K to 400~K, and found that  $\avg{\delta \omega}$ changed by approximately 30\% over this temperature range. It is evident from Figure \ref{fig:cfs_versus_energy} that the CFS at cryogenic temperatures is not much smaller than the CFS at room temperature. This fact implies that collisions could continue to limit the performance of next-generation cryogenic ion clocks, where the residual gas background is likely to be dominated by helium.

\section{Summary}
We have developed a non-perturbative framework for the detailed evaluation of collisional frequency shifts of atomic clocks. As a practical example of the method, we have estimated the shift of the 674 nm clock transition in Sr$^{+}$, where elastic collisions with helium are estimated to lead to a systematic fractional frequency shift of $-6(19) \times 10^{-21}$/nPa. Our framework can be easily applied to evaluate the collisional shifts of transitions used in other trapped-ion optical clocks.

\section*{Acknowledgments}
A.C.V is grateful to John Sipe for helpful discussions, and for sharing some unpublished notes at the outset of this investigation. We acknowledge support from NSERC, Canada Research Chairs, and a Society in Science - Branco Weiss Fellowship.

\bibliography{collisions}

\end{document}